\documentclass[prl,twocolumn,aps,showpacs,superscriptaddress]{revtex4}
\usepackage{graphicx}  % needed for figures
\usepackage{dcolumn}   % needed for some tables
\usepackage{bm}        % for math
\usepackage{amssymb}   % for math

\usepackage{amsmath}
\usepackage{braket}
\usepackage{color}

\begin{document}

\title{Nonlocal elasticity near jamming}

\author{Karsten Baumgarten}
\affiliation{Delft University of Technology, Process \& Energy Laboratory, Leeghwaterstraat 39, 2628 CB Delft, The Netherlands}

\author{Daniel V\aa gberg}
\affiliation{Delft University of Technology, Process \& Energy Laboratory, Leeghwaterstraat 39, 2628 CB Delft, The Netherlands}

\author{Brian P. Tighe}
\affiliation{Delft University of Technology, Process \& Energy Laboratory, Leeghwaterstraat 39, 2628 CB Delft, The Netherlands}

\date{\today}

\begin{abstract}
We demonstrate that the elasticity of jammed solids is nonlocal. By forcing frictionless soft sphere packings at varying wavelength, we directly access their transverse and longitudinal compliances without resorting to curve fitting. The observed wavelength dependence of the compliances is incompatible with classical (local) elasticity, and hence quantifies the amplitude of nonlocal effects. Three distinct length scales, two of which diverge, control the amplitude of both nonlocal effects and fluctuations about the mean response. Our results identify new, more accurate constitutive relations for weakly jammed solids, including emulsions, foams, and granulates.
\end{abstract}

\maketitle

\begin{figure}
\includegraphics[width=0.46\textwidth]{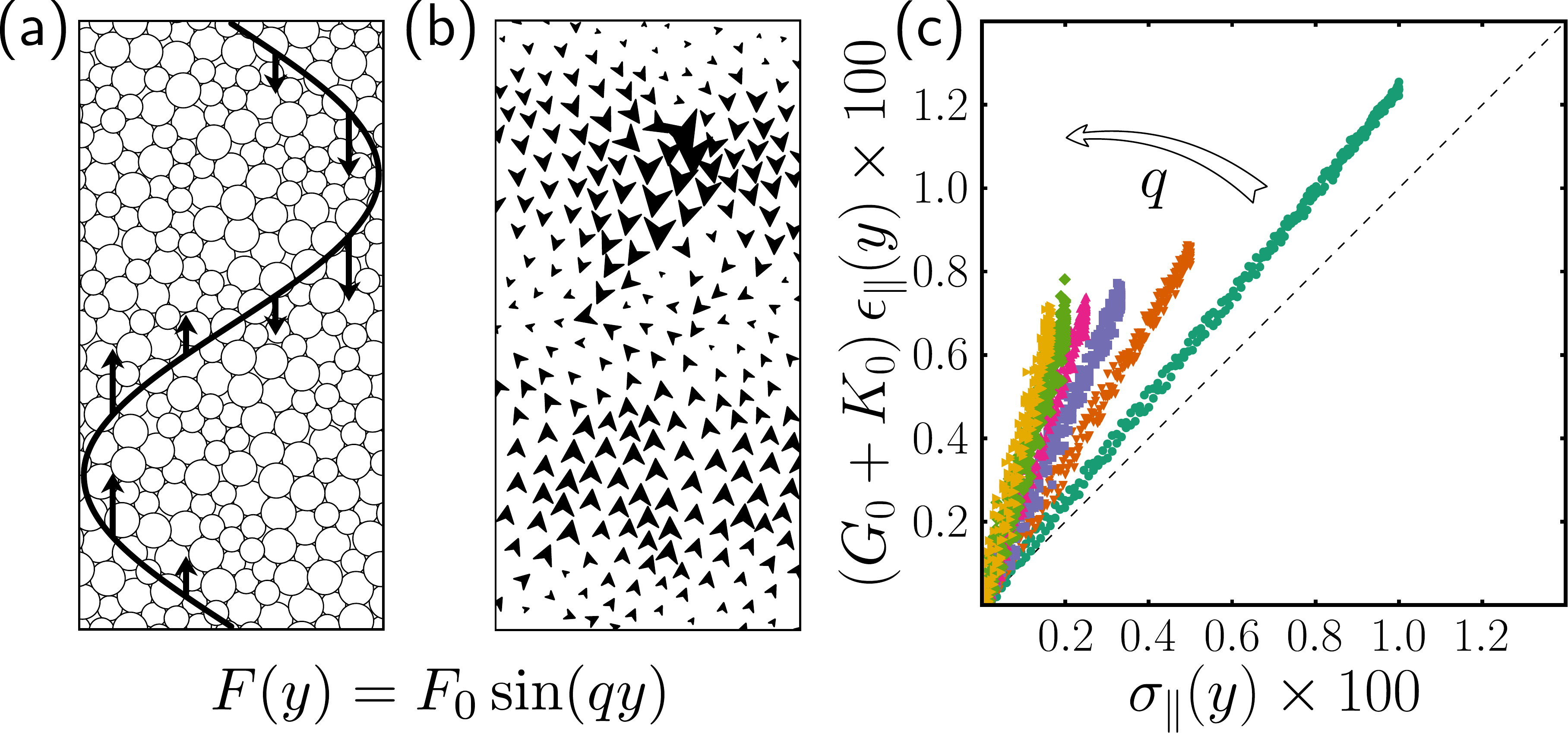}
\caption{(a) Sinusoidal forcing applied to a soft sphere packing, and (b) the resulting displacements. (c) A parametric plot of the constitutive relation is linear with a slope that increases with wavenumber $q$, violating classical elasticity (dashed line).
}
\label{fig:method}
\end{figure}

Classical linear elastic continuum theory 
is blind to structure: it contains no length scale(s) characteristic of, e.g., nearest neighbor spacing, interparticle interactions, or structural correlations \cite{landau}. 
As a result, continuum elasticity is valid only at asymptotically long wavelengths.
In practice, displacement fields can be accurate in ordered solids down to just a few nearest neighbor spacings, but deviations in amorphous materials are apparent over much longer distances \cite{tanguy02,maranganti07}. 

For example, Green's functions from molecular dynamics simulations of soft sphere packings -- a standard model for  emulsions, aqueous foams, and granular materials  \cite{ohern03,vanhecke10} -- depart significantly from elasticity when the packings are close to the (un)jamming transition \cite{ohern03,vanhecke10} at zero confining pressure $p$. 
Jammed solids are anomalously soft, with an abundance of low frequency vibrational modes and two associated diverging length scales $l^* \sim 1/p^{1/2}$ and $l_c \sim 1/p^{1/4}$, characteristic of longitudinal and transverse sound waves, respectively \cite{silbert05,wyart05}.
While details remain controversial, there is consensus that the break down of classical elasticity is governed by one or both of these length scales \cite{ellenbroek06,lerner14,karimi15}, both of which grow much larger than the mean particle size. 

Here we show for the first time that elastic constitutive relations near jamming are nonlocal \cite{bazant02,askes11}. 
Nonlocal constitutive relations are sensitive to spatial gradients; they ``know about'' microstructure via at least one length scale $\ell$, 
which appears as a crossover in moduli or compliances that vary with wavelength, unlike the constant moduli in classical elasticity.
By incorporating this wavelength dependence, nonlocal models push elasticity to shorter wavelengths while retaining the computational advantages of continuum methods. When the nonlocal length $\ell$ is large, classical elasticity breaks down rapidly and nonlocal descriptions become essential.

Our work is inspired by recent interest in nonlocal {\em rheology} near jamming \cite{pouliquen96,pouliquen01,komatsu01,aranson08,goyon08,pouliquen09,katgert10,nichol10,bouzid13,wandersman14,bouzid15b,kharel16,gueudre16}.  
Several nonlocal models have succeeded in predicting steady flow phenomena that previously defied description, such as flow below the nominal yield stress  \cite{goyon08,bocquet09} and wide shear bands in split-bottomed Couette cells \cite{kamrin12,henann13}. 
Yet, despite their successes (and a long history in engineering \cite{mindlin64,eringen83,bazant02,askes11}), nonlocal models have not been widely adopted -- at least in part because there are many competing variants, each with coefficients that must be determined empirically.

Here we identify nonlocal effects in soft sphere packings without assuming a particular model ahead of time or resorting to curve fitting. Our method is reminiscent of oscillatory rheology, which allows access to frequency-dependent moduli. By applying forcing that is periodic in space, rather than time, we measure wavelength-dependent compliances \cite{kuhn05,todd08}.
In addition to clear evidence of nonlocality near jamming, we find two diverging length scales, growing fluctuations, and surprising differences between compression and shear response.

\emph{Model system.---} We study mixtures of $N$ soft disks in $D=2$ dimensions with equal numbers of large and small disks having a 1.4:1 ratio of their radii, a commonly studied model system \cite{koeze16}. Unless noted otherwise, $N = 65,\!536$ prior to removing non-load bearing ``rattlers''. Contacting disks labeled $i$ and $j$ interact via a pair potential $V_{ij} = (1/2) k \delta_{ij}^2$, where $\delta_{ij}$ is the difference between the sum of the disks' radii and their center-to-center distance. Non-contacting disks do not interact. All results are reported in units where the spring constant $k$ and the small particle diameter $d$ are equal to 1. Packings are prepared in a bi-periodic $L \times L$ cell via instantaneous quench from infinite to zero temperature using a nonlinear conjugate gradient method \cite{vagberg11}, followed by a series of small volume changes to reach a target pressure. Particle displacements are determined by inverting $DN$ coupled linear equations involving the Hessian, the matrix of second derivatives of the potential energy with respect to the particle positions \cite{maloney06,tighe11}.
We employ the standard technique of ``removing the pre-stress'', which is equivalent to replacing each contact with a spring at its rest length \cite{ellenbroek09,schoenholz13}. Data with the pre-stress included are qualitatively similar but noisier.

\emph{Measuring nonlocal constitutive relations.---} 
We adapt a test that has been independently developed several times -- see Refs.~\cite{somfai05,karimi15} and especially \cite{kuhn05,todd08}, which explicitly make the connection to nonlocality. 
Packings are subjected to longitudinal and transverse force densities
\begin{align}
{\bf f}_\parallel(y) &= (0, f_\parallel)^T \sin{qy} 
\label{eqn:sinusoid1} \\
{\bf f}_\perp(y) &= (f_\perp, 0)^T \sin{qy} 
\label{eqn:sinusoid2}
\end{align}
with wavenumber $q$. 
These establish changes in the stress tensor with Fourier amplitudes $\delta \hat \sigma_{yy} (q) \equiv \hat \sigma_\parallel(q) = f_\parallel/q$ and $\delta \hat \sigma_{xy}(q) \equiv \hat \sigma_\perp(q) = f_\perp/q$, respectively.
We then measure the average displacement fields ${\bf u}_\parallel = (0, u_\parallel)^T$ and ${\bf u}_\perp = (u_\perp,0)^T$. Longitudinal forcing and response are illustrated in Fig.~1a and b.
We restrict ourselves to linear response \cite{vandeen14,boschan16}, though application to nonlinear response and flow is possible. 

In a classical and isotropic elastic continuum, a sinusoidal force density establishes a sinusoidal displacement field in phase with the forcing. Hence we can reproduce the constitutive relation by noting that a parametric plot of, e.g., the $y$-components of $q \, {\bf u}_\parallel(y)$ and $q^{-1} \, {\bf f}_\parallel (y)$ sweeps out the same curve as a conventional plot of strain $\epsilon_\parallel$ versus stress $\sigma_\parallel$. 
Classical elasticity predicts the curve will be linear (because we are probing linear response) with a constant slope $K_0 + G_0$ equal to the sum of the bulk and shear moduli, respectively, and independent of $q$ (because the theory is insensitive to strain gradients). 
In Fig.~1c we demonstrate that the second prediction fails near jamming:
the constitutive relation is indeed linear, but its slope varies with $q$ and approaches the classical prediction (dashed line) only as $q \rightarrow 0$, when spatial gradients are weakest. This is our first main result: the elasticity of jammed packings is indeed nonlocal. 

To quantify nonlocality, we measure the longitudinal compliance $\hat S_\parallel(q) = q^2 \hat u_\parallel(q) / f_\parallel$ and transverse compliance $\hat S_\perp(q) = q^2 \hat u_\perp(q) / f_\perp$ for each packing via direct Fourier transform of the displacement field. 
These two compliances fully determine the linear nonlocal constitutive relation \cite{eringen83}, which in Fourier space reads $\hat \sigma_{\alpha \beta}({\bf q}) = \hat C_{\alpha \beta \gamma \delta}({\bf q}) \, \hat \epsilon_{\gamma \delta}({\bf q})$ (summation implied). The tensor $\hat C$ has all the symmetries of the usual elastic coefficient tensor \cite{landau}, which has two independent elements in isotropic systems; these are fixed by $\hat S_\parallel = 1/\hat C_{yyyy}$ and $\hat S_\perp = 1/\hat C_{xyxy}$.

Local elasticity must be recovered for spatially uniform strains in translationally invariant systems. Hence
$\hat S_\parallel(0) = 1/[K_0 + G_0]$ and $\hat S_\perp(0) = 1/G_0$, where $K_0 \sim p^0$ and $G_0 \sim p^{1/2}$ obey known scaling relations near jamming \cite{vanhecke10}. 
Continuity of the $q = 0$ limit is not required, but will be verified numerically below.

\begin{figure}
\includegraphics[width=0.46\textwidth]{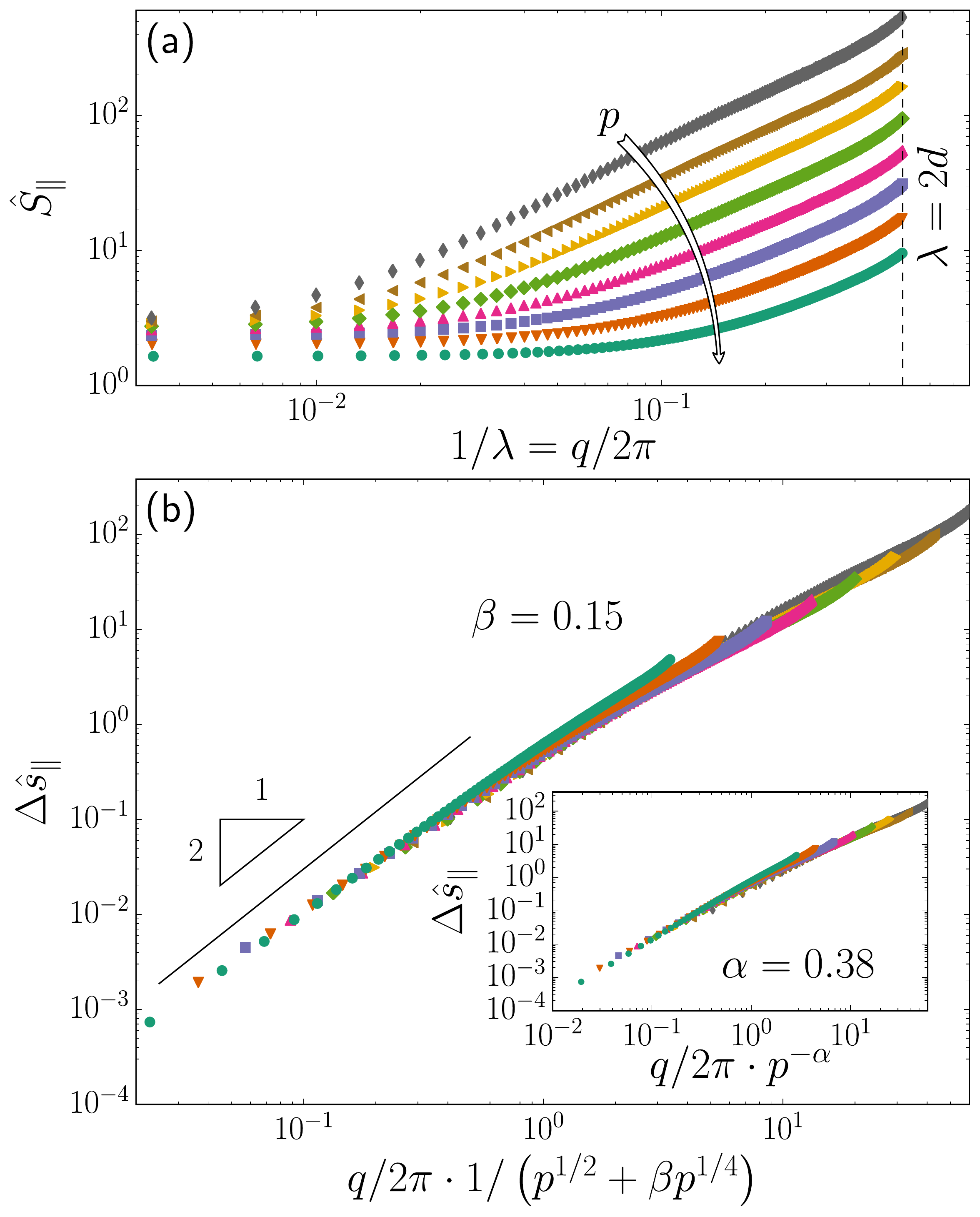}
\caption{
(a) Longitudinal compliance $\hat S_\parallel(q)$ versus inverse wavelength $\lambda$ for pressures $10^{-5.5} \le p \le10^{-2}$ in half decade steps.
(b) Data collapse of the excess compliance $\Delta \hat s_\parallel(q)$.
}
\label{fig:youngs_compliance}
\end{figure}

\emph{Longitudinal forcing.---} We first consider the response to longitudinal forcing. Fig.~2 depicts $S_\parallel(q)$ for a range of pressures close to jamming and wavenumbers $2\pi/L \le q \le \pi/d$. Data are averaged over approximately 1,000 configurations per condition.

Several features of $\hat S_\parallel(q)$ are noteworthy. First, each curve approaches a pressure-dependent plateau $\hat S_\parallel(0^+) $ as $q$ tends to zero. To determine whether the limit is continuous, we measure the local compliance $\hat S_\parallel(0)$ by subjecting each packing to a uniform stress in an independent test \cite{tighe11}. As shown in Fig.~2b,  the excess compliance $\Delta \hat s_\parallel(q) \equiv \langle  \hat S_\parallel(q) / \hat S_\parallel(0) \rangle - 1$ vanishes continuously with $q$, indicating a continuous limit.

The compliance shows a clear pressure-dependent crossover, which selects a nonlocal length scale $\ell_\parallel$. We now show that this length diverges with pressure. To do so we demonstrate that the excess compliance $\Delta \hat s_\parallel$  collapses to a master curve by when plotted versus the rescaled coordinate $q\ell_\parallel$. 
We first consider a simple power law ansatz $\ell_\parallel \sim 1/p^\alpha$, and obtain good data collapse for $\alpha = 0.38$ and pressures $10^{-5.5} \le p \le 10^{-2}$ (Fig.~2b, inset). However, we find that the value of $\alpha$ giving the best collapse decreases systematically as the highest pressures are removed from the dataset; e.g.~$\alpha \approx 0.33$ for $10^{-5.5} \le p \le 10^{-3}$. This shift indicates subdominant corrections to scaling are present. As noted above, there are two known diverging length scales near jamming; there is also evidence for an admixture of the two in the spatial structure of states of self stress \cite{sussman16}. 
By making a second ansatz $1/\ell_\parallel \sim 1/l^* + \beta/l_c \sim p^{1/2} + \beta \, p^{1/4}$, we obtain good collapse with $\beta \approx 0.15$. Unlike $\alpha$, removing higher pressures does not change our estimate of $\beta$. We therefore consider it likely that the longitudinal length $\ell_\parallel \rightarrow l_c$ at jamming. This is surprising insofar as $l_c$ is usually associated with shear \cite{silbert05}. We stress that a diverging nonlocal length implies sizable nonlocal corrections to classical elasticity, regardless of the precise value of the exponent.

\begin{figure}
\includegraphics[width=0.46\textwidth]{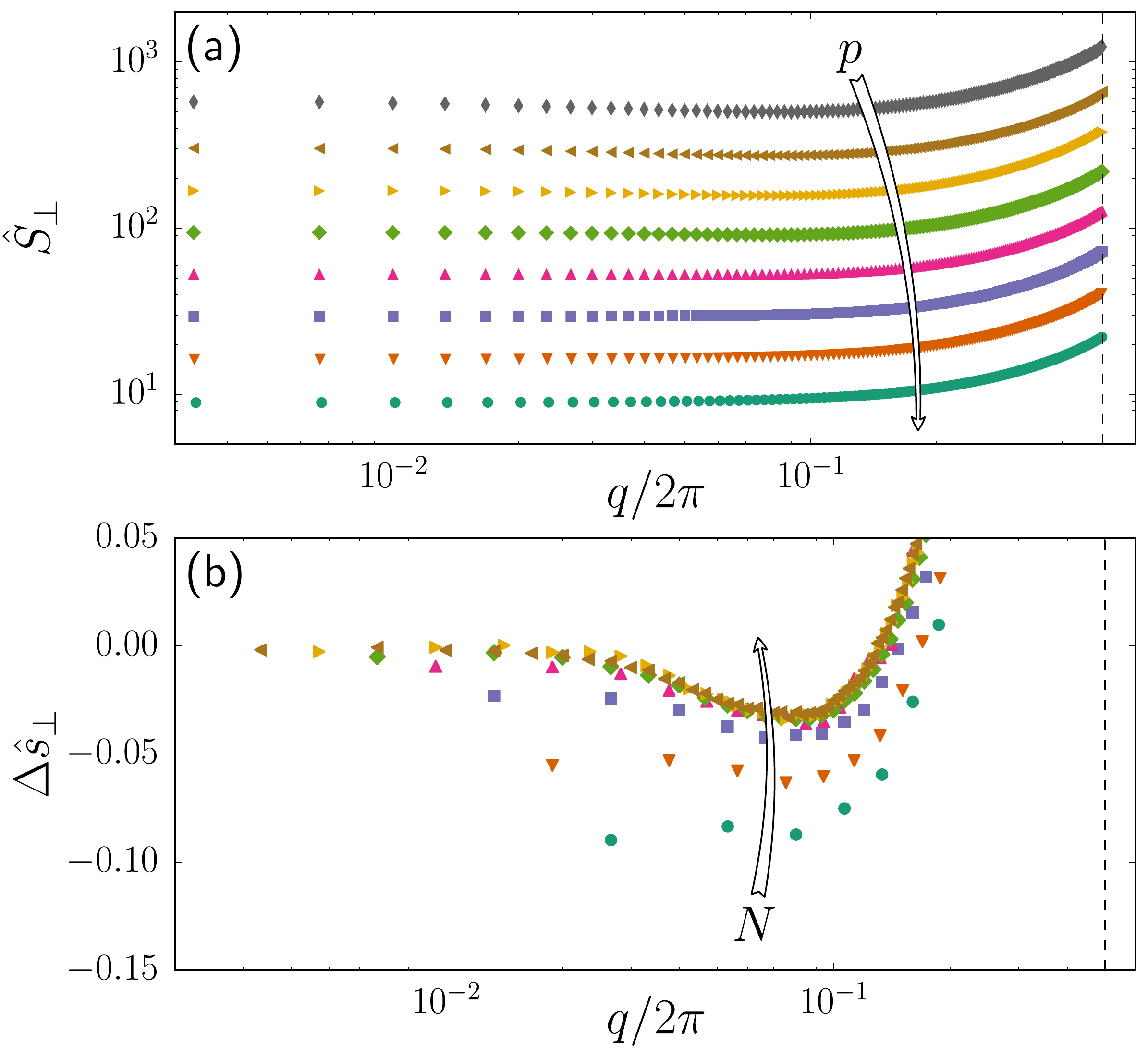}
\caption{(a) The transverse compliance shows no pressure-dependent crossover for $10^{-5.5} \le p \le 10^{-2}$. 
(b) The excess compliance is non-monotonic, with strong finite size effects. Shown here: $p = 10^{-4}$ and $2^{11} \le N \le  2^{16}$ in octave steps.
}
\label{fig:shear_compliance}
\end{figure}

\emph{Transverse forcing.---} 
Fig.~3 plots the transverse compliance $\hat S_\perp(q)$ for a range of pressures. While the general shape of the compliance curves echoes the longitudinal case, several differences stand out. First, the crossover scale $1/\ell_\perp$ is a constant on the order of the inverse particle size, independent of pressure. Hence the transverse length $\ell_\perp$ does not diverge near jamming, unlike $\ell_\parallel$. A similar $p$-independent crossover was noted in Ref.~\cite{karimi15} without making the connection to nonlocality.
The transverse compliance is non-monotonic, with an initial dip that appears to survive in the infinite system size limit (Fig.~3b). Despite the dip, the $q\rightarrow 0$ limit is again continuous, $\hat S_\perp(0^+) = \hat S_\perp(0)$. Finite size effects are stronger than in the longitudinal forcing case (not shown); they are also more dramatic than finite size effects under uniform strain \cite{goodrich14}, which can be neglected when $p \gg 1/N^2$ -- which holds for all data in Fig.~3.

\begin{figure}
\includegraphics[width=0.46\textwidth]{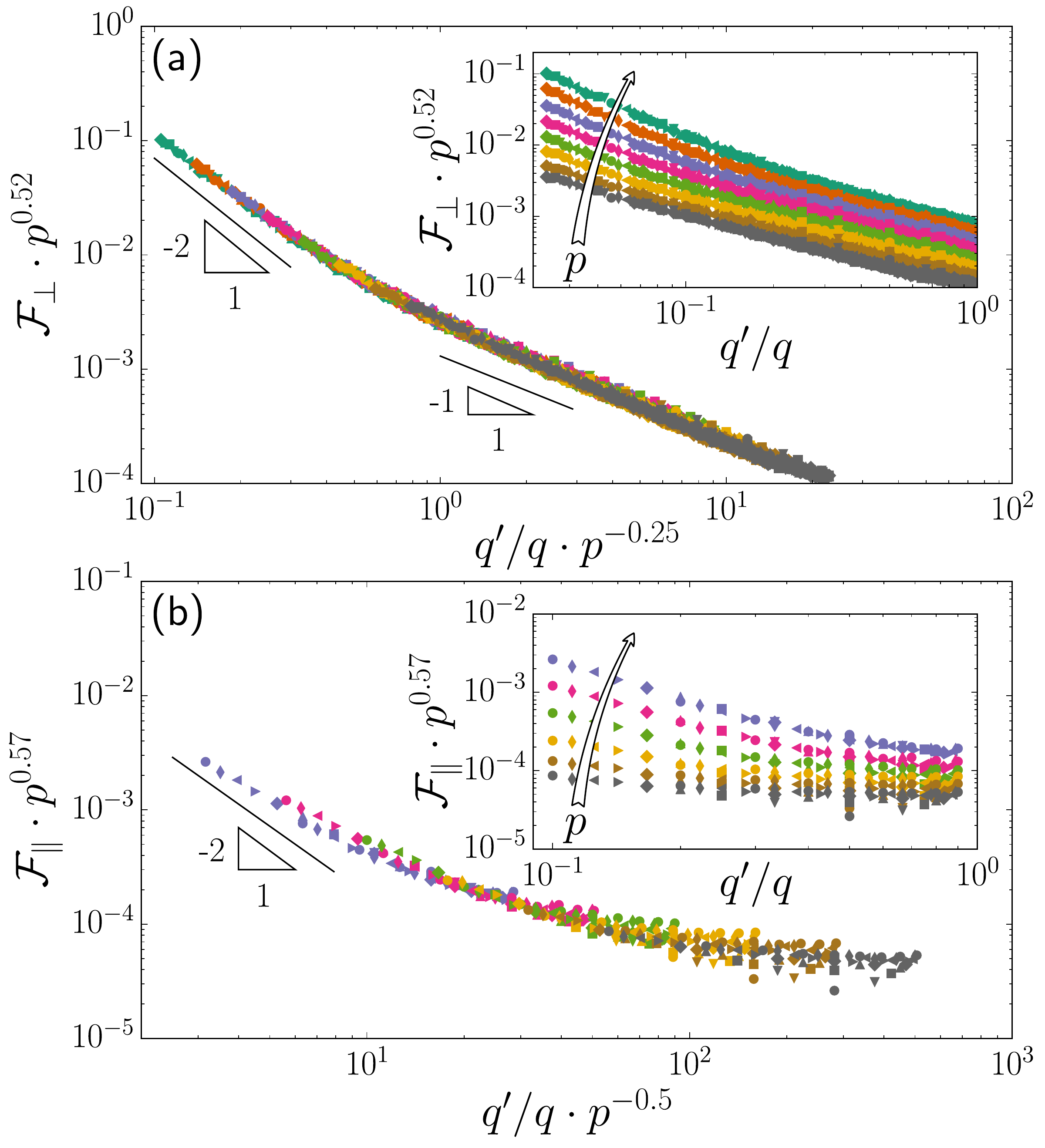}
\caption{
(a) Data collapse of the transverse Fourier spectrum before (inset) and after (main panel) rescaling $q'$ with $p^{-0.25}$.
(b) The longitudinal Fourier spectrum shows similar collapse before (inset) and after (main panel) rescaling $q'$ with $p^{-0.5}$.
}
\label{fig:fluctuations}
\end{figure}

\emph{Fluctuations.---}
It is apparent from Fig.~1b that individual particle displacements deviate from perfect sinusoidal response. These non-affine fluctuations can be quantified by the ensemble average of the ratio 
\begin{equation}
{\mathcal  F}_\circ(q'; q) = \left|\frac{\braket{u_\circ(q) | q_\circ'}}{\braket{u_\circ(q) | q_\circ}} \right| \,,
\end{equation}
where $\circ$ refers to $\parallel$ or $\perp$. ${\cal F}_\circ$ compares the projections of the $DN$-component displacement vector $\ket{u_\circ(q)} = \lbrace ({\bf u}_{\circ})_i \rbrace_{i = 1 \ldots N}$ on sinusoids with wavenumbers $q' \neq q$ and $q$.  The sinusoids' polarization matches the forcing.

We first consider transverse forcing. We restrict our focus to long wavelengths $q \le 30(2\pi/L)$, where $S_\perp$ is approximately flat, and consider only $q' < q$; these fluctuations have the largest amplitudes. Fig.~4a (inset) shows that for a given pressure, ${\cal F}_\perp$ collapses when plotted versus $q'/q$. The curves show a pressure-dependent crossover from steep to shallower decay. The data can be collapsed further still by plotting $p^{a_\perp} {\cal F}_\perp$ versus $(q'/q)/p^{1/4}$, with $a_\perp \approx 0.52$ (Fig.~4a, main panel). We conclude that transverse fluctuations are governed by the length scale $l_c$. 

Analyzing low-$q$ fluctuations under longitudinal forcing is more difficult due to the vanishing crossover near jamming. As a compromise we vary $q'$ and $q$ for $q' < q \le 10(2\pi/L)$, where the excess compliance $\Delta \hat s_\parallel$ is approximately quadratic for all accessed pressures. These fluctuations have a more complex dependence on $q$, as evidenced by slight but systematic spread in the data when plotted versus $q'/q$ -- see Fig.~4b (inset). Nevertheless, there is a clear $p$-dependent crossover, which can be collapsed by plotting $p^{a_\parallel} {\cal F}_\parallel$ versus $(q'/q)/p^{1/2}$, with $a_\parallel \approx 0.57$ (main panel). While the collapse is less convincing than ${\cal F}_\perp$, it suggests that longitudinal fluctuations are governed by the length scale $l^*$.

For both types of forcing, we observe data collapse only for sufficiently low $q'$. The restriction to $q' < q$ is strictly necessary in the longitudinal case; data fall off the master curve rapidly for larger $q'$. In the transverse case the fall off comes later and more gradually. We note that prior work has related $l^*$ \cite{ellenbroek06}, $l_c$ \cite{lerner14}, or both \cite{karimi15} to (deviations from) classical elastic Green's functions \cite{ellenbroek06,lerner14,karimi15}.

{\em Discussion.---} 
We have demonstrated that discrete, finite-ranged interactions between soft spheres near jamming give rise to  continuum constitutive relations that are nonlocal. Nonlocal effects are stronger in deformations involving compression, as reflected in the distinct length scales $\ell_\perp$ and $\ell_\parallel$; the former remains finite, while the latter diverges at the jamming transition. Fluctuations about the mean nonlocal response are governed by the diverging length scales $l^*$ and $l_c$ in longitudinal and transverse response, respectively. 

For analytical modeling, it is often desirable to assign a functional form to the compliances. 
Noting that symmetry requires $\hat S_\parallel$ and $\hat S_\perp$ to be even functions of $q$ in isotropic materials, one anticipates the leading term in an expansion of $\Delta \hat s_\circ$ to be quadratic in $q$, as verified in Fig.~2b. Truncating the expansion leads to the following constitutive relations (in scalar form for simplicity):  
\begin{align}
(1 - \ell_{\perp}^2 \, \partial^2) \sigma_\perp
&= 2G_0 \, \epsilon_\perp  
\label{eqn:quad2} \\ 
(1 - \ell_{\parallel}^2 \, \partial^2) \sigma_\parallel 
&= (K_0 + G_0) \, \epsilon_\parallel  \,.
\label{eqn:quad1} 
\end{align}
Eq.~(\ref{eqn:quad2}) provides a good description of the transverse response over a wide range of $q$; note the minus sign neglects the dip in $\hat S_\perp$.
Eq.~(\ref{eqn:quad1}) is a significant improvement over its local counterpart, though it misses the slow bending over of $\Delta \hat s_\parallel$ apparent in Fig.~2b.

Of course one would like to have an accurate description of the nonlocal compliances over the whole range of $q$. Fitting functions are an option, though they lack physical insight. Micromechanical models such as effective medium theory (EMT) would be preferable. While we expect that EMT can predict the nonlocal transverse compliance, it fails to capture the longitudinal compliance even for spatially uniform forcing \cite{ellenbroek09b}.

The sinusoidal forcing technique used here is in no way restricted to soft spheres -- it can be used to test for nonlocal effects in a wide range of materials. It is straightforward to implement numerically and can also be implemented in experimental systems that allow for forcing in the bulk, such as thermoresponsive microgels and granular monolayers. In the jamming context, obvious extensions include acoustic dispersion relations \cite{schoenholz13}, nonlinear forcing \cite{boschan16}, and steady flow \cite{olsson07,tighe10c}.

{\em Acknowledgments.---} We thank Wouter Ellenbroek and Edan Lerner for helpful discussions.
We acknowledge financial support from the Netherlands Organization for Scientific Research (NWO), and the use of supercomputer facilities sponsored by NWO Physical Sciences.

\bibliographystyle{apsrev}
% \bibliography{tighe}

\end{document}